\documentclass[twocolumn,showpacs,preprintnumbers,amsmath,amssymb]{revtex4}
\usepackage{graphicx}
\usepackage{dcolumn}
\usepackage{bm}
\begin{document}
\title{Quantum information processing using Josephson junctions coupled
through cavities}
\author{Shi-Liang Zhu$^{1,2}$}
\author{Z. D. Wang$^{1,3}$}
\author{Kaiyu Yang$^1$}
\affiliation{$^1$Department of Physics, University of Hong Kong,
Pokfulam Road, Hong Kong, China\\ $^2$Department of Physics, South
China Normal University, Guangzhou, China\\ $^3$Department of
Material Science and Engineering, University of Science and
Technology of China, Hefei, China}
\begin{abstract}
Josephson junctions have been shown to be a promising solid-state
system for implementation of quantum computation. The two-qubit
gates are generally realized by the capacitive coupling between
the nearest neighbour qubits. We propose an effective Hamiltonian
to describe charge qubits coupled through the microwave cavity. We
find that nontrivial two-qubit gates may be achieved by this
coupling. The ability to interconvert localized charge qubits and
flying qubits in the proposed scheme implies that quantum network
can be constructed using this large scalable solid-state system.
\end{abstract}
\maketitle

Quantum information processing (QIP) with a large number of qubits
is now attracting increasing interest. So far, a number of systems
have been proposed as potentially viable qubit models. Among a
variety of qubits implemented, solid-state qubits are of
particular interest because of their potential suitability for
integrated devices. Charge qubits based on Josephson junctions
have been shown to be a promising solid-state candidate for
implementation of quantum computation (QC)
\cite{Shnirman,Averin,Makhlin,Nakamura,Falci,Zhu_prl2002,Wang,Pashkin}.
QIP tasks usually involve not only computation but also
communication. However, whether Josephson junctions are also
suitable for quantum communication is still an important open
question.

The basic criteria for QIP have been described in Refs.
\cite{DiVincenzo,Preskill}. Among them, realization of a
universal set of quantum
gates play a central role in QC. Besides, that the gates
can act on any pair of qubits is also a necessary
element for fault tolerant computation\cite{Preskill}.
Moreover,
the ability to interconvert stationary and flying qubits,
and to faithfully transmit flying qubits between specified
nodes are also required
for quantum communication\cite{DiVincenzo}.

In this paper, we show that a new system consisting of Josephson
junctions coupled through microwave cavities fulfills the above
requirements, and thus is a promising candidate for QIP. This
system possesses at least three distinctive merits  . (i) A
serious limitation of solid-state computers is that the
decoherence time in these systems is relatively short. However,
from the report in a recent experiment, it is possible that
quantum coherence of a large number qubits may be easier to
maintain if junctions locate within a high quality microwave
cavity\cite{Barbara}. (ii) The nontrivial two-qubit gate acting on
any pair of qubits can be realized, and possible fault tolerant
geometric quantum computation\cite{Falci,Zhu_prl2002,Wang}
proposed in the absent of cavity is still workable. Thus the
combination of different fault tolerant approaches is possible,
and may be helpful for overcoming the infamous decoherence
effects.
Here we provide a new experimentally feasible method to realize
two-qubit gates: coupling charge qubits through a high quality
cavity, and show that the combination of capacitive as well as
cavity coupling in symmetric SQUID or just the cavity coupling in
asymmetric SQUID allow the implementation of two-qubit gates
between any pair of qubits. (iii) The present scheme is able to
interconvert stationary charge qubits and flying photon qubits,
and faithfully transmit flying qubits between specified nodes in
quantum network. The ideal quantum transmission between charge
qubits in different cavities can be achieved by cavity QED
techniques\cite{Cirac}. Thus the quantum network based on solid
state quantum computers may be connected by using transmission
fibre, and photons as flying qubits in the scheme clearly
represent the best qubit carrier for fast and reliable
communication over long distances.

The single Josephson junction qubit we
considered is shown in Fig. 1(a) \cite{Makhlin}.
It consists of a small superconducting box with n excess
Cooper-pair charges,
formed by an
SQUID with capacitances $C_{Jm}$ $(m=1,2)$ and
Josephson coupling energies $E_{Jm}$,
pieced by a magnetic flux $\phi$. A control gate voltage $V_g$ is
connected to the system via a gate capacitor $C_g$.
The Hamiltonian of the system becomes
\begin{equation}  \label{H_single}
H=E_{ch} (n-\bar{n})^2-E_{J1} cos\gamma_1-E_{J2} cos\gamma_2,
\end{equation}
where $n$ is the number operator of (excess)
Cooper-pair charges on the box,
$E_{ch}=2e^2/(C_g+C_{J1}+C_{J2})$ is
the charging energy, $\bar{n}=C_g V_g/2$
is the induced charge and can be controlled
by changing $V_g$. $\gamma_m$ is
the gauge-invariant phase difference between
points on opposite sides of the
$m$th junction. Assuming that the Josephson
junction locates within a single-mode
resonant cavity, then
$\gamma_m=\varphi_m-\frac{2\pi}{\phi_0}\int_{l_m}
{\bf A}_m \cdot d {\bf l}_m$, where $\varphi_m$ is the
phase difference (of the superconducting wave function)
across the $m$th junction in a particular gauge,
and may take the same value
$\varphi$ \cite{Almaas}.
${\bf A}_m$ is the vector potential in the same gauge, and the
line integral is taken across the $m$th junction, and along the arrows in
Fig. 1(a). ${\bf A}_m$ may be divided into two parts
${\bf A}_m^\prime+{\bf A}_m^\phi $, where the first term arises from the
electromagnetic field of the cavity normal mode
(which can be described as an oscillator)
and the second term arises
from the magnetic flux $\phi$. In the Coulomb gauge,
${\bf A}_m^\prime $ takes the form
$\sqrt{\hbar/2 \omega V}(a+a^{\dagger})\hat{\epsilon}$\cite{Almaas},
where $\hat\epsilon$ is the
unit polarization vector of the cavity mode, $V$ is the volume of the
cavity,
$a$ and $a^\dagger$ are the annihilation and creation operators for the
quantum oscillators, and $\omega$ is its frequency.
Therefore, we have
\begin{equation}
\label{Phase_difference}
\frac{2\pi}{\phi_0}\int_{l_m} {\bf A}_m\cdot d {\bf l}_m
=\frac{2\pi }{\phi_0} \int_{l_m}
{\bf A}^\phi_m\cdot d {\bf l}_m + g(a+a^{\dagger}).
\end{equation}
where $g=2e \hat{\epsilon} \cdot {\bf l}/\sqrt{2\varepsilon \omega V\hbar}$
is the
coupling constant between the junctions and the cavity,
with $l$ the thickness of the insulating layer in the junction. For simplicity, we
assume that $g_m=g$. As for ${\bf A}^\phi$, we have another constraint: $%
\oint_C {\bf A}^\phi d {\bf l} =\phi$, where the integral path $C$ is along
the dashed line in Fig. 1(a).

We consider systems in the charging regime where
$E_{ch} \gg E_{Jm}$,
then a convenient basis is formed
by the charge states, parameterized by the number of Cooper pairs $n$ on the
box, and $\varphi$ is its conjugate $n=-i\hbar\partial/\partial(\varphi)$.
They satisfy the standard commutation relation: $[\varphi, n]=i$. In this
basis the Hamiltonian (\ref{H_single}) reads
\begin{eqnarray}  \label{H_coupling}
H = &&\sum_n [E_{ch} (n-\bar{n})^2 |n\rangle \langle n|  \nonumber \\
&& -\frac{E_J (\phi)}{2}(e^{-i[g (a+a^\dagger)+\beta]} |n+1\rangle \langle n
| +H.c.)],
\end{eqnarray}
where
\begin{eqnarray}
\label{E_J}
&& tan\beta = \frac{E_{J1}-E_{J2}}{E_{J1}+E_{J2}}tan(\frac{\pi \phi}{\phi_0}),\\
\label{Beta}
&&  E_J (\phi) = \sqrt{(E_{J1}-E_{J2})^2+4 E_{J1}E_{J2} cos^2 (\pi\phi/\phi_0)}
\end{eqnarray}
with $\phi_0=\pi\hbar/e$ being the flux quantum.
At temperature much lower
than the charging energy and the gate voltage
tuning close to a degeneracy ($\bar{n}\sim 1/2$),
the relevant physics is captured by considering only the
two charge eigenstates $n=0,1$, which constitute the basis
$\{|0\rangle,|1\rangle\}$ of the computation
Hilbert space of the qubit.

\begin{figure}[htbp]
\centering
\includegraphics[width=8cm]{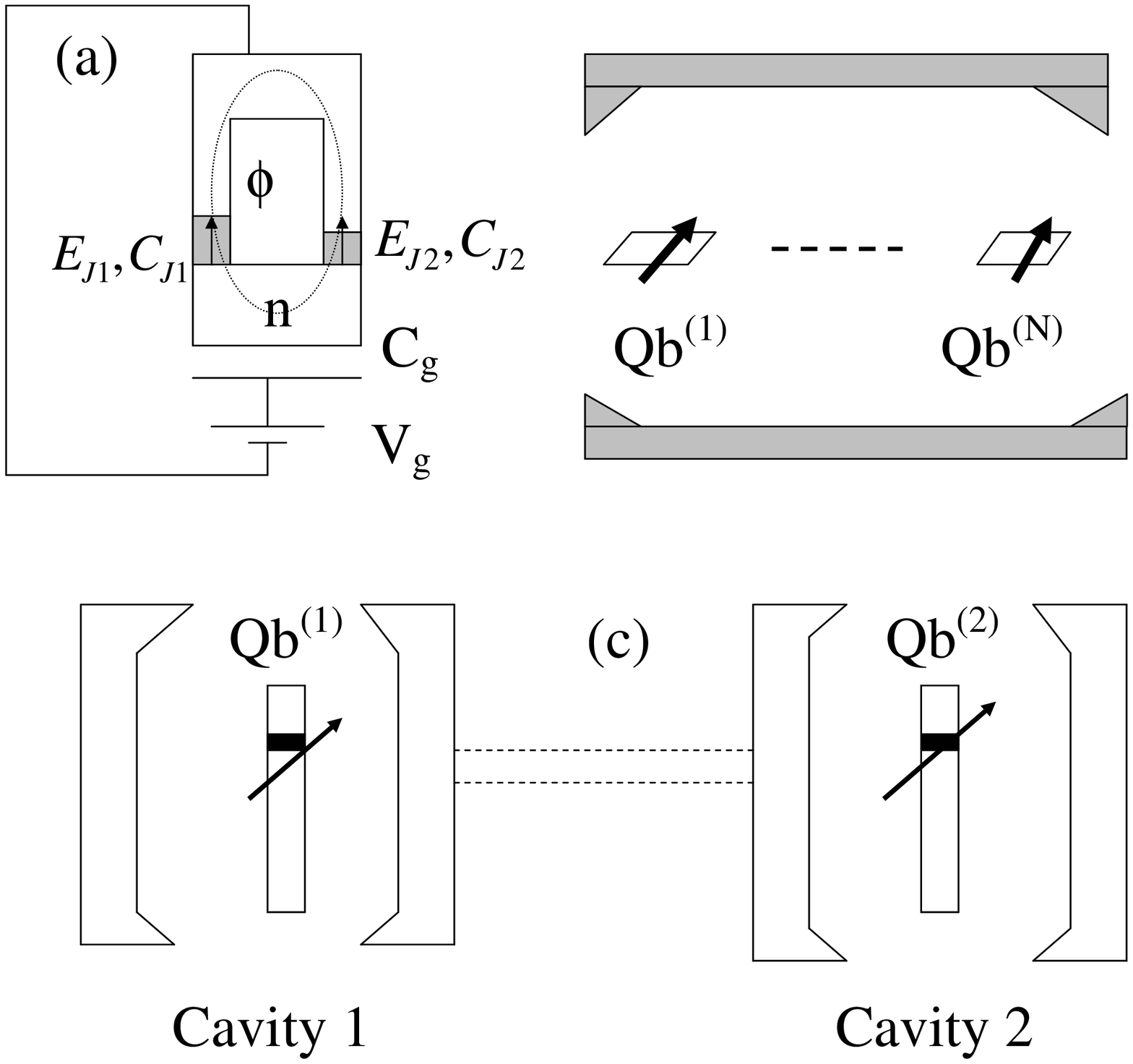}
\vspace{-3.5cm} \caption{Josephson qubit systems. (a) A single
Josephson qubit. (b) Josephson qubits in a cavity. (c) Josephson
qubits in cavities connected by transmission fibre.} \label{fig}
\end{figure}

If we have $N$ such qubits located within
a single-mode cavity [Fig.1(b)],
to a good approximation, the total system can be
considered as $N$ two-state
systems coupled to a quantum harmonic oscillator\cite{Almaas}.
In this case,
the system considered here can be described by the Hamiltonian
$H = H_0+H_{int}$, where
\begin{eqnarray}
\label{H_0}
H_0 &=& \hbar \nu (a^\dagger a+\frac{1}{2}) + \sum_{j}^{N} E_{\bar{n}_k}
\sigma_j^z, \\
\label{H_int}
H_{int} &=& -\frac{1}{2}\sum_{j}^{N} E_J(\phi_j) (e^{-i[g
(a+a^\dagger)+\beta_j]} \sigma_j^{+}+H.c)
\end{eqnarray}
with $E_{\bar{n}_k}=E_{ch}(\bar{n}_k-1/2)$. A spin notation is used for the
qubit $j$ with Pauli matrices
$\{ \sigma^{x}_j,\sigma^{y}_j,\sigma^{z}_j \}$,
and $\sigma_j^{\pm}=(\sigma^x_j \pm i\sigma^y_j)/2$. For simplicity, we
have assumed the same $E_{ch}$, $E_{J1}$ and $E_{J2}$  for all
different qubits. The tunable parameters $E_J(\phi_j)$
and $\beta_j$ have the same forms as those in
Eqs. (\ref{E_J}) and (\ref{Beta}),
where $\phi_j$ is the magnetic flux pieced the $j$th Josephson
charge qubit. It is remarkable that the main
parameters $E_{\bar{n}_k}$ and $E_J(\phi_j)$
in the Hamiltonian can be controlled independently for every
qubit. Furthermore, Eq.(\ref{H_int})
representing the interaction between
charge qubits and cavity QED is essential
in the implementation of QIP.

We now present two examples to demonstrate that QC may be
accomplished by using the above Josephson junction system. For
universal QC, we need to realize only two kinds of noncommutable
single-qubit gates and one nontrivial two-qubit gate\cite{Lloyd}.

The first example is QC using asymmetric
SQUID loop ($E_{J1}\neq E_{J2}$).
The single-qubit gates can be realized
when a qubit energy gap is far from
the cavity energy, thus the qubit is decoupled
from the cavity. In this
case the effective Hamiltonian for qubit $k$ reads\cite{Makhlin}
\begin{equation}
H_k=E_{\bar{n}_k}\sigma_k^z - E_J(\phi_k) (\sigma_k^x
cos\beta_k-\sigma_k^y sin\beta_k).
\end{equation}
When both $\bar{n}_k$ and $\phi_k$ are time-independent, the evolution
operator is obtained explicitly
\begin{equation}  \label{U_single}
U(\gamma_k)=\exp(-\frac{i}{\hbar}\int_0^t H_k dt)
=\exp(-i\gamma_k \sigma_{{\bf n}} {\bf \cdot n}),
\end{equation}
where
${\bf n}=(-E_J(\phi_k) \cos\beta_k, E_J(\phi_k)
\sin\beta_k, E_{\bar{n}_k} )/E_k$
with
$E_k=\sqrt{E_J^2(\phi_k)+E_{\bar{n}_k}^2}$,
$\gamma_k=E_k t/\hbar$ and $\sigma_{{\bf n}}$
is Pauli matrix along  the direction
${\bf n}$. We may check that $U({\bf n}_1)$
and $U({\bf n}_2)$
are noncommutable if ${\bf n}_1\ne \pm {\bf n}_2$.
Consequently, the
universal single-qubit gates can be realized by suitable choosing
$\bar{n}_k$ and $\phi_k$.

We now address that nontrivial two-qubit gate may be achieved by
the cavity coupling. In the condition that $g\sqrt{n+1}$ is well
below unity (the Lamb-Dicke limit), we may expand Eq.
(\ref{H_int}) in powers of $g$ and neglecting rapidly rotating
terms. By choosing the first blue sideband frequency
($E_{\bar{n}_{k}}=\hbar \nu $), we find a transformation
$$
R_{k}^{+}(\theta _{k},\beta _{k})=\exp \left[ -i\frac{\theta _{k}}{2}
(ie^{-i\beta _{k}}\sigma _{k}^{+}a+H.c.)\right]
$$
with $\theta _{k}=E_J (\phi_k) gt/\hbar$.
Similarly, by choosing
the first red sideband frequency
($E_{\bar{n}_{k}}=-\hbar \nu$), we have another transformation
$$
R_{k}^{-}(\theta _{k},\beta _{k})=\exp \left[ -i\frac{\theta _{k}}{2}%
(ie^{-i\beta _{k}}\sigma _{k}^{+}a^{\dagger }+ H.c.)\right].
$$
A controlled-NOT gate for control qubit $j$ and target qubit $k$
can be realized by using $R_{k}^{\pm}$ and single
qubit rotations, for
example
\begin{eqnarray}  \label{CNOT}
U_{jk}^{CNOT} = && Z_{j}(-\pi /(2\sqrt{2}))R_{j}^{-}(\pi ,\beta_j
)H_{k}P_{k}  \nonumber \\
&&Z_{k}(-\pi /(2\sqrt{2}))H_{k}R_{j}^{-}(\pi ,\beta_j )
\end{eqnarray}
for any value of $\beta_j $ \cite{Childs}.
Here $Z_{j}(\zeta )$ is a phase
gate for qubit $j$, $H_{k}$ is the Hadamard gate and
$P_{k}=R_{k}^{+}(-\pi/2,0)R_{k}^{+}(-\pi \sqrt{2},
-\pi /2)R_{k}^{+}(\pi /2,0)$. The gates
described by Eqs. (\ref{U_single}) and (\ref{CNOT})
consist of a universal
set of quantum gates using charge qubits in asymmetric SQUID loop. The
requirement for fault tolerant computation is fulfilled, as $j$ and $k$ can
be any pair of qubits.

The second example is QC using symmetric SQUID loop
($E_{Jm}=E_{J0})$.
The Hamiltonian is given by $H=H_0+H_1+H_2$, where
\begin{eqnarray}
\label{H_1}
&& H_{1} = -\frac{1}{2}\sum_{j}^{N} E_J^0 (\phi_j)(e^{-ig (a+a^\dagger)}
\sigma_j^{+}+H.c.),\\
&& \label{H_2}
H_{2} = E_c\sum_{\langle i,j \rangle}(\bar{n}_i-n_i)(\bar{n}_j-n_j)
\end{eqnarray}
with $E_J^0 (\phi_j)=2E_{J0} \cos(\pi\phi/\phi_0)$. Here $H_0$ is the
same as that of Eq. (\ref{H_0}), and $H_2$ with $\langle i,j\rangle$
denoting the nearest
neighbor qubits represents the capacitive couplings between
qubits\cite{Falci}.
We also consider this coupling,
because it is unlikely to work
out easily a nontrivial two-qubit gate by
using only the coupling with cavity.

The single qubit gates may be realized when $H_2$ is set to zero. In the
rotationed frame
$U_0 (t)= \exp[-i\nu t(a^{\dagger}a+1/2)]\exp(-i E_{\bar{n}_k} t
\sigma_j^z)$, the interaction Hamiltonian is given by $H_{int}^{\prime}
= U_0^\dagger H_{int} U_0 \approx H_a+H_b$, where
\begin{eqnarray}
\label{H_a}
H_a &=& -\sum_j^{N} E_J^0 (\phi_j)\sigma_j^{x}, \ \ \ (E_{\bar{n}_k}=0)
\\
\label{H_b}
H_{b} &=& \frac{1}{2}\sum_{j}^{N} E_J^0 (\phi_j)
(ig a \sigma_j^{+}+H.c.). \
\ \ (E_{\bar{n}_k}=\nu)
\end{eqnarray}
Thus from Eq. (\ref{H_a})
we have a unitary operator
$U_x (\gamma^x_k)=\exp[-i\gamma^x_k\sigma_k^x/2]$
with $\gamma^x_k=2E_J^0 (\phi_k)t/\hbar$
for the qubit $k$ by
choosing $E_{\bar{n}_k}=0$.
On the other hand, it is seen from Eq. (\ref{H_b})
that the interaction between the
cavity and qubit $k$ is decoupled by choosing
$\phi_k=(i_k+1/2)\phi_0$ with
$i_k$ an integer, and the evolution operator is
derived as $U_z (\gamma^z_k)=\exp[-i\gamma^z_k\sigma_k^z/2]$
with $\gamma^z_k=2E_{\bar{n}_k} t/\hbar$. The
gates described by $U_x (\gamma^x_k)$ and $U_z (\gamma^z_k)$ are a
well-known universal set of single-qubit gates.

Also by choosing $\phi_k=(i_k+1/2)\phi_0$, the Hamiltonian of two
qubits becomes $
H=E_{\bar{n}_1}\sigma_1^z+E_{\bar{n}_2}\sigma_2^z+
E_c(\bar{n}_1-n_1)(\bar{n}_2-n_2)$. Then we find a conditional
phase gate in computational basis given by
\begin{equation}  \label{C_phase}
U=\mbox{diag} \left(
e^{i\gamma_{00}},e^{i\gamma_{01}},e^{i\gamma_{10}},e^{i\gamma_{11}}
\right),
\end{equation}
where $\gamma_{{n_1}{n_2}}=-\omega_{{n_1}{n_2}}t$ with
$\hbar\omega_{{n_1}{n_2}}$ the eigenenergy of state
$|{n_1}{n_2}\rangle$, it is nontrivial under the condition
$\gamma_{00}+\gamma_{11} \neq \gamma_{01}+\gamma_{10}$ (mod
$2\pi$). A similar gate was addressed in Ref.
\cite{Falci,Zhu_prl2002}. The gates $U_x(\gamma^x_k)$,
$U_z(\gamma^z_k)$ and (\ref{C_phase}) consist of a universal set
of quantum gates using charge qubits in symmetric SQUID loop.

It is remarkable that the previously proposed geometric quantum
gates\cite{Falci,Zhu_prl2002,Wang,Zhu_pra2003} are still workable
in the above two example, as the Hamiltonians are essentially in
the same forms. Thus an intrinsically fault tolerant QC is
possible in the present systems. On the other hand, the scheme
based on symmetric SQUID loop has some special advantages compared
with that using asymmetric SQUID: first, the coupling between the
cavity and charge qubits may be experimentally tunable to zero
(but can not touch zero for asymmetric case). Thus the two-qubit
gates may be accomplished by using the same approach as that in
Ref. \cite{Falci,Zhu_prl2002}. Second, the eigenstates of the
Hamiltonian may be tuned to degenerate. Then the relative phase of
two logical states is zero during idle periods, but the
nondegenerate feature in asymmetric SQUID loop requires that the
phase difference induced by the energy spacing between logical
states must be controlled with high accuracy\cite{Makhlin}.

Another main advantage for coupling by cavities
is on the quantum communication. Since
swapping gates are essential on this direction,
we now address three very
useful kinds of restricted swapping gates
based on the cavity QED technique
in the symmetric SQUID. Note that
a slight modification of the approach can also be applicable in
the nonsymmetric SQUID.

If we consider a fixed qubit $k$ and pursue the evolution of the system
followed by Eq. (\ref{H_b}) for a certain time $t$, we obtain an evolution
operator
\begin{equation}
U_{kp}(\Gamma_k t)=\exp [ -i\Gamma_k t (i\sigma_k^\dagger
a +H.c.)]
\end{equation}
with $\Gamma_k=g E_J^0 (\phi_k)/2\hbar$.
This
transformation keeps the state $|0_k\rangle
|0\rangle_{ph}$ unaltered, whereas
\begin{eqnarray*}
|0_k \rangle|1\rangle_{ph} &\rightarrow& \cos(\Gamma_k t)|0_k
\rangle|1\rangle_{ph} +\sin(\Gamma_k t)|1_k \rangle|0\rangle_{ph}, \\
|1_k \rangle|0\rangle_{ph} &\rightarrow& \cos(\Gamma_k t)|1_k
\rangle|0\rangle_{ph} -\sin(\Gamma_k t)|0_k \rangle|1\rangle_{ph}
\end{eqnarray*}
with the subscript 'k' ('ph') denoting the $k$th qubit (photon in the
cavity). Then we have the following
transformation
\begin{equation}
\label{Swapping_kp}
(\alpha |0_k \rangle+\beta |1_k \rangle ) |0\rangle_{ph}
\stackrel{U_{kp}^\prime}{\longrightarrow}
|0_k\rangle (\alpha |0 \rangle_{ph}+\beta |1 \rangle_{ph} )
\end{equation}
between qubit $k$ and photon in the cavity, where $\alpha$ and $\beta$ are
complex numbers. Here $U_{kp}^\prime$ is a short denotation
of $U_{kp}[(2n-1/2)\pi]$ with $n$ an integer. Gate (\ref{Swapping_kp})
is the basis for interconverting stationary and flying qubits.
Moreover, we find that a swapping
gate between qubits $k$ and $j$ in the same cavity is given by the
operator $U_{jp}^\prime U_{kp}^\prime$, i.e.,
$$
(\alpha |0_k \rangle+\beta |1_k \rangle ) |0_j\rangle |0\rangle_{ph}
\stackrel{U_{jp}^\prime U_{kp}^\prime}{\longrightarrow}
|0_k\rangle (\alpha |0_j
\rangle+\beta |1_j \rangle ) |0\rangle_{ph}.
$$
It is worth pointing out that $j$ and $k$ may be any two qubits in a same
cavity, then this gate
can help to realize any two-qubit gate acting on any pair of qubits as long
as it can be achieved on two fixed qubits. Thus this swapping gate is very
useful to overcome the drawback arise from that the two-qubit gates
described by Eq. (\ref{C_phase}) act only on the nearest neighbor qubits.

Moreover, the ideal quantum transmission\cite{Cirac} (swapping gate)
between two cavities $1$ and $2$ (see Fig. 1(c))
\begin{eqnarray}
\nonumber
&& (\alpha |0_1 \rangle+\beta |1_1 \rangle )
|0_2\rangle \otimes|0_{1}\rangle_{ph}
|0_{2}\rangle_{ph} |\mbox{vac}\rangle \\
\label{Swapping_12}
\longrightarrow
&& |0_1\rangle (\alpha |0_2 \rangle+\beta |1_2 \rangle )
\otimes|0_{1}\rangle_{ph} |0_{2}\rangle_{ph} |\mbox{vac}\rangle
\end{eqnarray}
with $|\mbox{vac}\rangle$ the vacuum state of
the free electromagnetic modes connecting the cavities,
can be accomplished by appropriately selecting the
controllable parameters
$\bar{n}_i$ and $\phi_i$ in each cavity.
Following the approach described in Ref. \cite{Cirac},
we find that the
evolution equations to achieve the ideal transmission in Eq.
(\ref{Swapping_12}) are given by
\begin{equation}
\label{Evolution}
\dot{\alpha}_i=g\beta_i E_J^i/2,\ \ \dot{\beta}_i=-g\alpha_i
E_J^i/2-\kappa\beta_1,\ \ (i=1,2),
\end{equation}
where $\kappa$ is the loss rate of each cavity,
$\{\alpha_1,\alpha_2,\beta_1,\beta_2 \}$
are the expansion coefficients of the wave function
$|\Psi (t)\rangle$ in the basis
$\{ |1_1 0_2 \rangle |0_{1} 0_{2}\rangle_{ph},  |0_1
1_2 \rangle |0_{1} 0_{2}\rangle_{ph},
|0_1 0_2 \rangle |1_{1} 0_{2}\rangle_{ph},
|0_1 0_2 \rangle |0_{1} 1_{2}\rangle_{ph} \}$,
and we have chosen ${\bar{n}}_i$
so that $E_{\bar{n}_i}=\nu/2$. The mathematical problem is
now to find $\phi_i (t)$ such that $\alpha_1 (-\infty)
=\alpha_2 (+\infty)=1$ and
Eq. (\ref{Evolution}) are fulfilled. A type of symmetric
solutions [$E^0_J(\phi_2(t))=E^0_J(\phi_1(-t))$]
can be found by the approach outlined in
Ref.\cite{Cirac}. For example, we find that
\begin{eqnarray*}
&&\phi_1=(\phi_0/\pi) \arccos (\kappa/gE_{J0}),\\
&&\phi_2=(\phi_0/\pi) \arccos[\kappa e^{-\kappa t/2}
\cos(\sqrt{3}\kappa t/2-\pi/3)/g\alpha_2 E_{J0}],
\end{eqnarray*}
where $\alpha_2=\sqrt{1-e^{-\kappa t}[1+\cos(\sqrt{3}\kappa t
-\pi/6)/\sqrt{3}]/2}$ $(t\ge 0)$, is a set of appropriate
analytical solutions. Therefore the ideal quantum transmission
between two nodes of a quantum network may be accomplished using
microwave photons in this system\cite{Cirac}.

The quantum network proposed here consists of spatially separated
nodes connected by quantum communication channels. Each node is a
quantum computer using Josephson junctions, which is able to
store quantum information in quantum bits and processes this
information locally using quantum gates. The transmission between
the nodes of the network is accomplished using microwave photons
with the cavity QED technique.
Furthermore, once a simpler quantum network is accomplished,
one of important applications is to test
Bell's inequality in this mesoscopic system\cite{He}.


To conclude, we have presented a new approach to couple Josephson
junction qubits and have shown that this system satisfies all the
acknowledged theoretical criteria for the construction of quantum
information network. Nevertheless, it is a big challenge to
implement this kind of network experimentally.

This work was supported by the RGC grant of
Hong Kong under No. HKU7114/02P
and a URC fund of HKU.
S.L.Z was supported in part by SRF for ROCS, SEM,
the NSF of Guangdong under Grant No. 021088,
and the NNSF of China under Grant No. 10204008.

\end{document}